\documentclass[5p]{elsarticle}
\usepackage{lineno,hyperref}
\usepackage{graphicx}
\usepackage{url} 
\usepackage{dblfloatfix}  

\journal{Journal of \LaTeX\ Templates}
\usepackage{color}
\biboptions{authoryear}

\begin{document}

\begin{frontmatter}

\title{New members of Datura family}

\author[mymainaddress]{Rosaev A.}
\ead{hegem@mail.ru}

\author[mysecondaryaddress]{Pl\'avalov\'a E.}
\ead{plavalova@komplet.sk}

\address[mymainaddress]{NPC Nedra, Yaroslavl, Russia}
\address[mysecondaryaddress]{Astronomical Institute, Slovak Academy of Science, Slovakia}
\begin{abstract}
Asteroid families are groups of minor planets which have a common origin in catastrophic disruptions. Young asteroid families are very interesting because they represent the product of their parent body's fragmentation before orbital and physical evolutionary processes could have changed them. A group of minor asteroids associated with the largest body Datura(1270) is of particular interest because it has enough known members and resides in the inner part of the main asteroid belt and is easy to observe. Up to now, 7 members of this family are known. Here we discuss three new members of the Datura Family: 338309(2002~VR17), 2002~RH291 and 2014~OE206. To prove that these recently-discovered members belong to the Datura family, we conducted numeric integration with all gravitational perturbation over the last 800~kyrs. In the results, we have found that 338309(2002~VR17) and 2002~RH291 are very close to the mean orbit of this family throughout the calculation. In the case of 2014~OE206, it has a strongly chaotic orbit. The possible explanation of this is in the resonance character of its orbit.
\end{abstract}
\end{frontmatter}

\section{Introduction}
The observations of the main asteroid belt have brought the discoveries of asteroid families – groups of fragments from collisions of the large asteroids. Up to now, more than 120 asteroid families have been discovered \cite{2015arXiv150201628N}. Most of the large asteroid families are at least 1~Gyrs old. The dynamical history of old asteroid families is complicated by a number of limiting factors such as the incomplete data of the influence of the Yarkovsky effect, chaos effect, subsequent collisions, etc. \citep{1999Icar..142...63M,2005Icar..179...63B,2006AJ....132.1950N}. The younger an asteroid family is, the more straightforward it is to analyse because they represent the product of their parent body fragmentation before orbital and physical evolutionary processes have had change them. 

\section{Datura family} 
A group of minor planets associated with the largest body Datura(1270), is of particular interest because it has enough known members and resides in the inner part of the main asteroid belt, which makes observation clearer. The Datura family was discovered by \cite{2006Sci...312.1490N}, after which, both its dynamical evolution and physical properties were rigorously studied. The model of rotational fission systems is described by \citep{2010Natur.466.1085P}. According to their results, if the mass ratio of fragments is significant, the rotational period of the primary grows longer. The very long period of rotation and the very elongated shape in the case of 2003 CL5(90265) \citep{2009A&A...507..495V}, may be indirect evidence of its recent breakup and may highlight its possible mechanism.
The Datura family consist of one large 10 km-sized asteroid (parent body) and a few small minor planets; possible fragments of a catastrophic breakup. Up to now only 7 members of the Datura family were known \citep{2015arXiv150201628N,2006Sci...312.1490N,2009A&A...507..495V}, all of them belonging to the so-called S-type of minor planets \cite{1984PhDT.........3T}. 

\begin{table*}[t]
\caption{Osculating orbital elements of the Datura family members at epoch 16-01-2009 (JD 2454848).}
\begin{center}
\begin{tabular}{l l r r r r r} 
\hline 
 Number& \multicolumn{1}{c}{Object} & \multicolumn{1}{c}{$\omega$} & \multicolumn{1}{c}{$\Omega$} & \multicolumn{1}{c}{\textit{i}} & \multicolumn{1}{c}{\textit{e}} & \multicolumn{1}{c}{\textit{a}}  \\
\hline \hline  
1270 & Datura & 258.84072020 & 97.88191931 & 5.98954802 & 0.207917356 & 2.234316251 \\
60151 & 1999 UZ6 & 260.57151084 & 96.79987218 & 5.99341398 & 0.207812911 & 2.235186383 \\
89309 & 2001 VN36 & 266.85629891 & 92.976202113 & 6.02164403 & 0.20634112 & 2.23559504 \\
90265 & 2003 CL5 & 261.84375257 & 95.69777810 & 5.99543662 & 0.207471114 & 2.234865786 \\
203370 & 2001 WY35 & 260.44606900 & 96.87210762 & 5.99154981 & 0.207435802 & 2.235226671 \\
215619 & 2003 SQ168 & 259.39568050 & 97.46790499 & 5.99082425 & 0.207908027 & 2.234266272 \\
\textbf{338309} & \textbf{2002 VR17} & \textbf{260.60046232} & \textbf{96.80636247} & \textbf{5.98940220} & \textbf{0.207747446} &
\textbf{2.235169351} \\
 & 2003 UD112 & 263.12446374 & 95.47877895 & 6.00471739 & 0.206934529 &
2.234566113 \\
   & \textbf{2002 RH291} & \textbf{262.04302362} & \textbf{95.75928088} & \textbf{5.99579553} & \textbf{0.207613218} & \textbf{2.235326612} \\
  & \textbf{2014 OE206} & \textbf{261.72701226} & \textbf{96.26530222} & \textbf{6.00041657} & \textbf{0.206975298} & \textbf{2.235606600} \\
\hline 
\end{tabular}
\end{center}
\footnotesize{NOTE: The three new members of the Datura family are shown in bold.}
\label{elements}
\end{table*}
\section{The three new members of the Datura family}

As it is noted by \cite{2006Sci...312.1490N}, the discovery of new members of the Datura family is of great interest in order to study their group dynamics. Here we report three new members of the Datura family. In order to find new members of the Datura family, we studied 634109 orbits of asteroids from the Lowell Observatory catalogue \citep{1994IAUS..160..477B}. We extracted the bodies which possess the values of the orbital elements in intervals of  $2.234<a<2.236$, $0.2 <e<0.21$, $5.9<i<6.1$ according to the values of the orbital elements of 1270~Datura. We tested these bodies and several of their clones by numerical integration. 

In our research, we have found three new members of the Datura family in addition to those listed in \cite{2015arXiv150201628N,2006Sci...312.1490N,2009A&A...507..495V}. These three new members are \textit{338309~2002~VR17}, \textit{2002~RH291} and \textit{2014~OE206} \citep{2015EPSC...10..434R}. Table~\ref{elements}. shows the osculating elements for both the seven old members and the three new shown in bold. These new members of the Datura family are faint and small: as expected by absolute magnitudes, they are from 80 (2014~OE206) up to 500 meters (2002~RH291 and 338309) in diameter, making them comparable with known member 203370 (Tab.~\ref{obsparamet}). It is easy to see, that the orbits of the two new members 338309(2002~VR17) and 2002 RH291 are determined much better than the orbit of 2003UD112, which was initially included in the Datura family \citep{2006Sci...312.1490N}. And the precision of the orbit of 2014~OE206 is comparable to 2003UD112 (Tab.~\ref{obsparamet}).

\begin{table*}[t]
\caption{Some observational parameters of the new members of the Datura family in comparison with the originally known member 2003~UD112.}
\begin{center}
\begin{tabular}{l l r r r } 
\hline 
 Number & \multicolumn{1}{c}{Object}   & \multicolumn{1}{c}{H} &   \multicolumn{1}{c}{Arc [days]}  \\
\hline \hline  
1270 & Datura & 12.5 & 37118.1\\
60151 & 1999 UZ6 & 16.1 & 6308.05\\
89309 & 2001 VN36 & 16.2 & 7915.94\\
90265 & 2003 CL5 & 15.9 & 7245.02\\
203370 & 2001 WY35 & 17.1 & 6828.77\\
215619 & 2003 SQ168 & 17.0 & 4702.05\\
338309 & {2002 VR17} & 17.7 & 5794.84 \\
 & 2003UD112 & 17.9 & 25\\
   & 2002 RH291 & 17.9 & 4743.76 \\
  & 2014 OE206 & 19.3 &  26 \\
\hline 
\end{tabular}
\end{center}
\footnotesize{NOTE: Some observational parameters of the new members of the Datura family. Where; $H$ is the absolute magnitude, $Arc$ is the orbital arc in days, spanned by observations used in an orbital computation by {\cite{1994IAUS..160..477B}}.
}
\label{obsparamet}
\end{table*}
\begin{figure}
   \centering 
     \includegraphics[width=8.5cm]{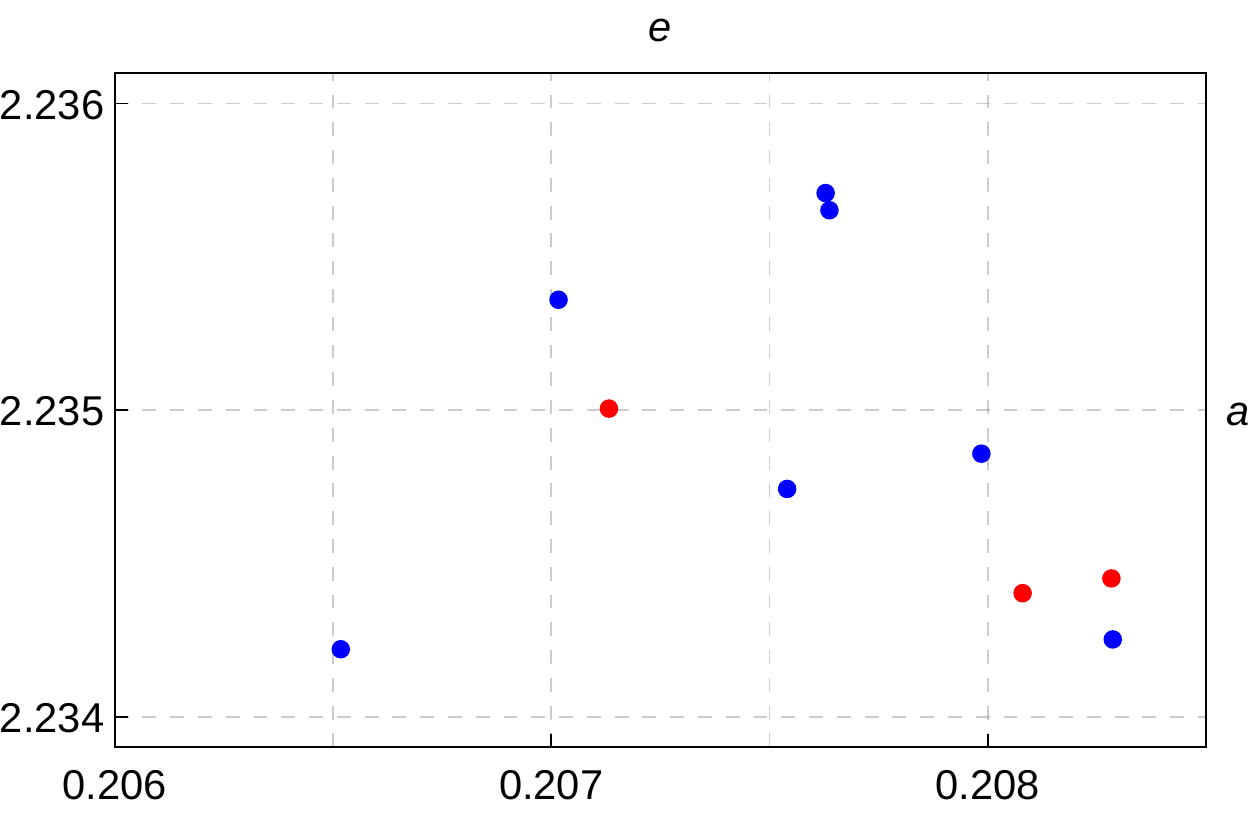}
     \includegraphics[width=8.5cm]{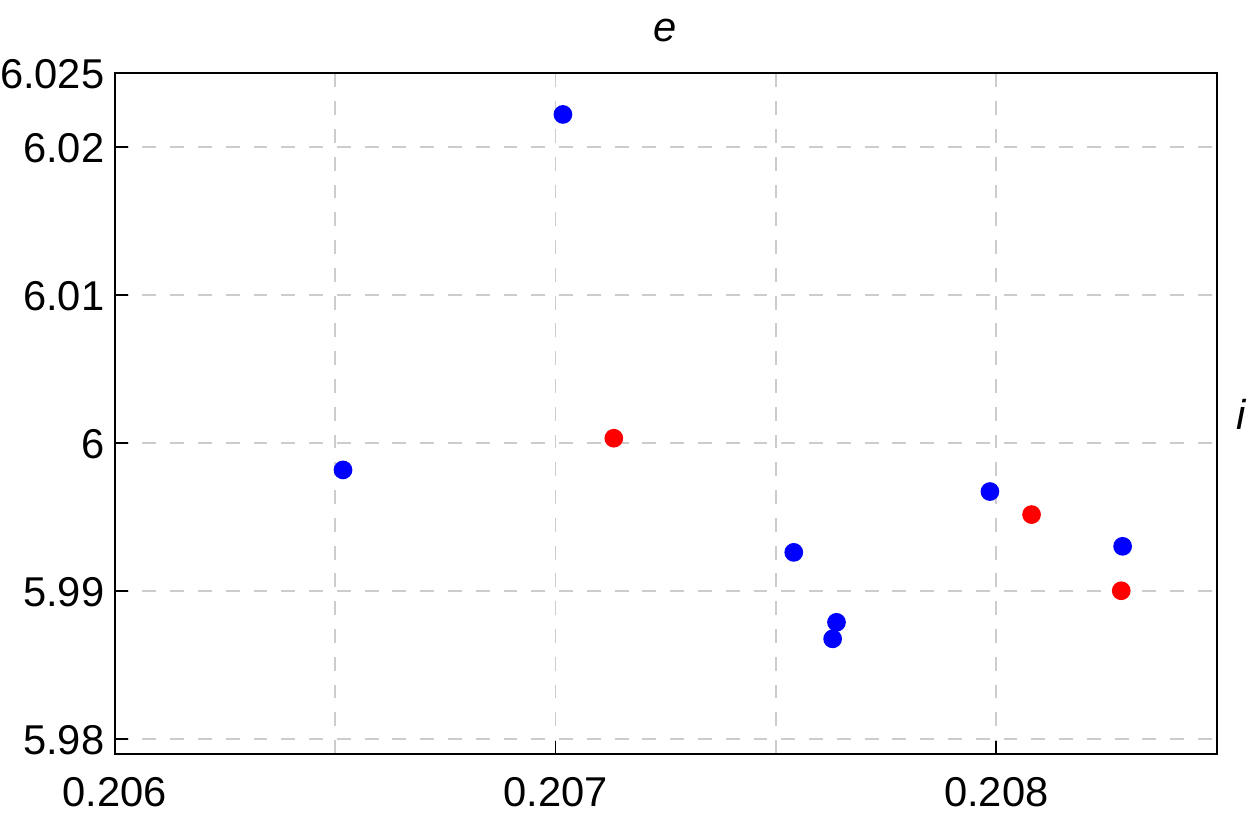}
     \caption{The positions of the Datura family members in $a, e$ (top figure) and $i,e$ (botom figure) coordinates. The new members are shown as red dots and the original members as blue dots.}
\label{aeposition}
 \end{figure}
\begin{figure}
   \centering 
     \includegraphics[width=8.5cm]{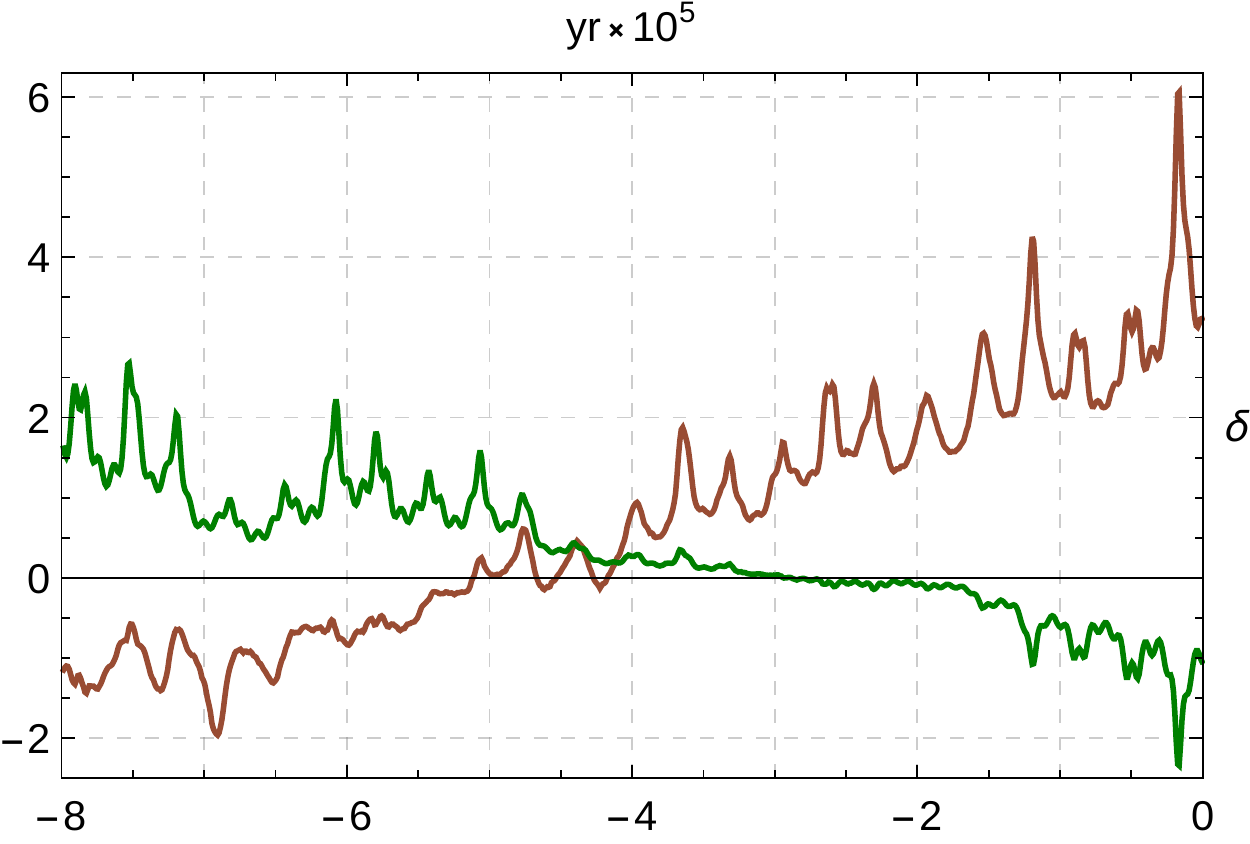}
     \caption{The average of differences in the mean longitude of the ascending nodes $\Omega$ for 2002~RH291 (brown line) and for 2002~VR17(338309) (green line); both belonging to 1270~Datura. }
\label{2002RH291}
 \end{figure}

\begin{figure}
   \centering 
     \includegraphics[width=8.9cm]{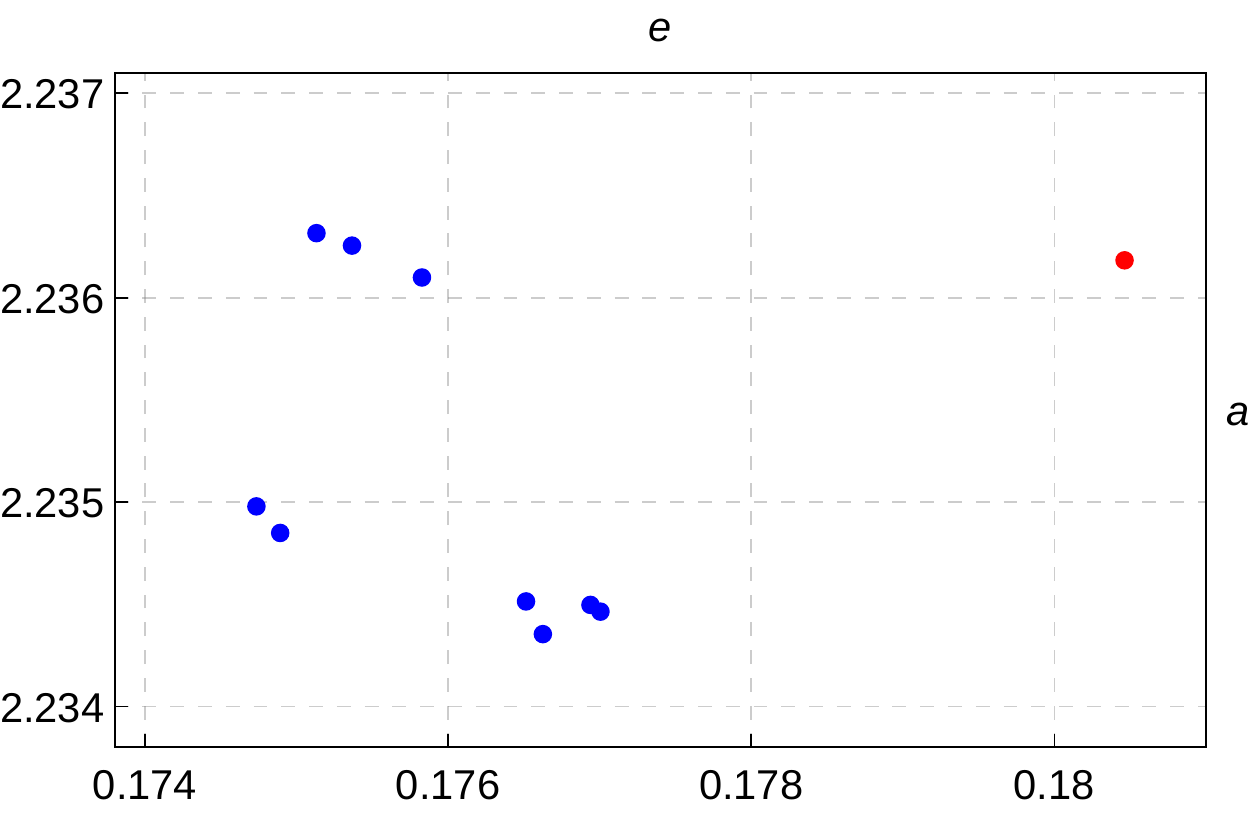}
      \caption{The positions of the Datura family members in $a$ and $e$ coordinates 300 kyr ago. Red point - 2014 OE206.}
\label{2014OE206}
 \end{figure}

To prove that these new members belong to the Datura family, we studied their numerical integration. The equations of the motion of the systems were numerically integrated 800~kyrs into the past, using the N-body integrator Mercury \citep{1999MNRAS.304..793C} and the Everhart integration method \citep{1985dcto.proc..185E}. The initial state vectors for planets were obtained from JPL DE405 ephemerides at the same time as the elements of the Datura-group asteroids \citep{1995JPL..314....S}. In addition, we use very few clones to establish the initial element's uncertainties for each of the Datura family’s new member's orbital evolutions. The distribution of Datura family members in the coordinates $(a, e)$ and and $(i,e)$\footnote{Throughout the article we have used standard notations for orbital elements $a$ -- semi-major axis in au, $e$ -- eccentricity, $i$ -- inclination, $\Omega$ -- longitude of ascending node, $\omega$ -- perihelion argument, $\pi$ -- longitude of perihelion (the angular elements are in degrees).} is given in Figure~\ref{aeposition}, where the new members are shown as red dots and the original members as blue dots. All the asteroids of the Datura family, including new members, form a compact group in the orbital element's phase space, true at least for 2002~VR17 and 2002~RH291 throughout all considered time intervals. In Figure~\ref{2002RH291}, the evolution of the angular elements for the two new members of the family are given, relative to 1270~Datura's angular elements. It is evident that the values of the angular elements of the new members are close to the largest member of the family in all considered time intervals. For other orbital elements we have similar relations. The orbital elements also remain similar throughout the whole interval of integration (Fig.~\ref{2014OE206}).
The age of the Datura family is estimated to be $450\pm50$~kyrs old according to \cite{2009A&A...507..495V} and $530\pm20$~kyrs old according to \citep{2006Sci...312.1490N}; that is to say that a question still lingers over the most precise age estimate. This inconsistency could mean that the question of the Datura family's age is still open.

\begin{table*}[t]
\caption{Proper orbital elements of the original and new members of Datura family and the averaged over 300 kyrs integration semi-major axis($a$)}
\begin{center}
\begin{tabular}{l l r | r r r } 
\hline 
 &  & &\multicolumn{3}{c}{Synthetic proper elements} \\
 Number & \multicolumn{1}{c}{Object}   & \multicolumn{1}{c}{Average a [au]} & \multicolumn{1}{|c}{$a$ [au]} &  \multicolumn{1}{c}{$e$} & \multicolumn{1}{c}{$\sin (i)$}  \\
\hline \hline  
1270 & Datura & 2.234694 & 2.23469& 0.153413 & 0.092307 \\
60151 & 1999 UZ6 & 2.234867 & 2.23489 & 0.153397 & 0.092331 \\
89309 & 2001 VN36 & 2.235201 & 2.23584 & 0.153883 &  0.092586 \\
90265 & 2003 CL5 & 2.234943 & 2.23494 & 0.153505 & 0.092397 \\
203370 & 2001 WY35 & 2.234865 & 2.23488 & 0.153482 & 0.092397 \\
215619 & 2003 SQ168 & 2.234663 & 2.23465 & 0.153450 & 0.092318 \\
338309 & 2002 VR17 & 2.234959 & 2.23494 & 0.153503 & 0.092313\\
 & 2002 RH291 & 2.235002 & 2.23490&  0.153533 & 0.092316 \\
\hline 
\end{tabular}
\end{center}
\label{synteticpropelements}
\end{table*}

As it seems in Figure~\ref{2002RH291}, the orbital evolution of the new members of the Datura family is not in conflict with the estimated age of \cite{2009A&A...507..495V} being $450\pm50$~kyrs. The argument of the perihelion is somewhat better than the node longitude for the determination of age. In the case of 2014~OE206, we have a strongly chaotic orbit. According to our results, the semi-major axis of 2014~OE206 significantly increased about $50$~kyrs ago (Fig. \ref{2014OE206}). This effect presents a small variation of the initial data of this asteroid in the range of $2.23485<a<2.2360$ au for the semi-major axis and at least $36.79^\circ<M<36.81^\circ$ for the range of the mean anomaly. In fact, the orbit of 2014~OE206 is a typical example of so-called stable chaotic orbit \citep{1997Icar..125...13M}.

The possible explanation of this is in the resonance character of its orbit. As it was adopted by \cite{2006Sci...312.1490N}, the orbit of 2001~VN36(89309) was assumed to be too uncertain because of resonance-related chaoticity (mean motion resonance $9:16$ to Mars). Except noted above, there is a commensurability with a resonance of $7:2$ to Jupiter in the vicinity of the Datura family, but its effect is weaker. The initial mean anomalies of 2001~VN36(89309) and 2014~OE206 are very close, so we can suppose a finite range in the mean anomaly of the instability of the orbit caused by resonance. Resonance is also presented in the orbital evolution of the other members of the family but in a weaker form. The variations of the initial values of other orbital elements (eccentricity, inclination, pericenter  and node longitude) have a relatively small effect on the orbital evolution of 2014~OE206.

Elementary consideration has shown that there has been no encounter with Mars closer than $0.01$ au. However, the change of semi-major axis in the case of 2014 OE206 is significant; $50$~kyrs ago, this asteroid leaped from one side of $9:16$ resonance to another. On the other hand, its other orbital elements remained very close to the mean elements of the other members of the Datura family. So, we can conclude that 2014~OE206 has belonged to the Datura family for at least the last 50 kyrs.

\begin{figure}
   \centering 
     \includegraphics[width=8.5cm]{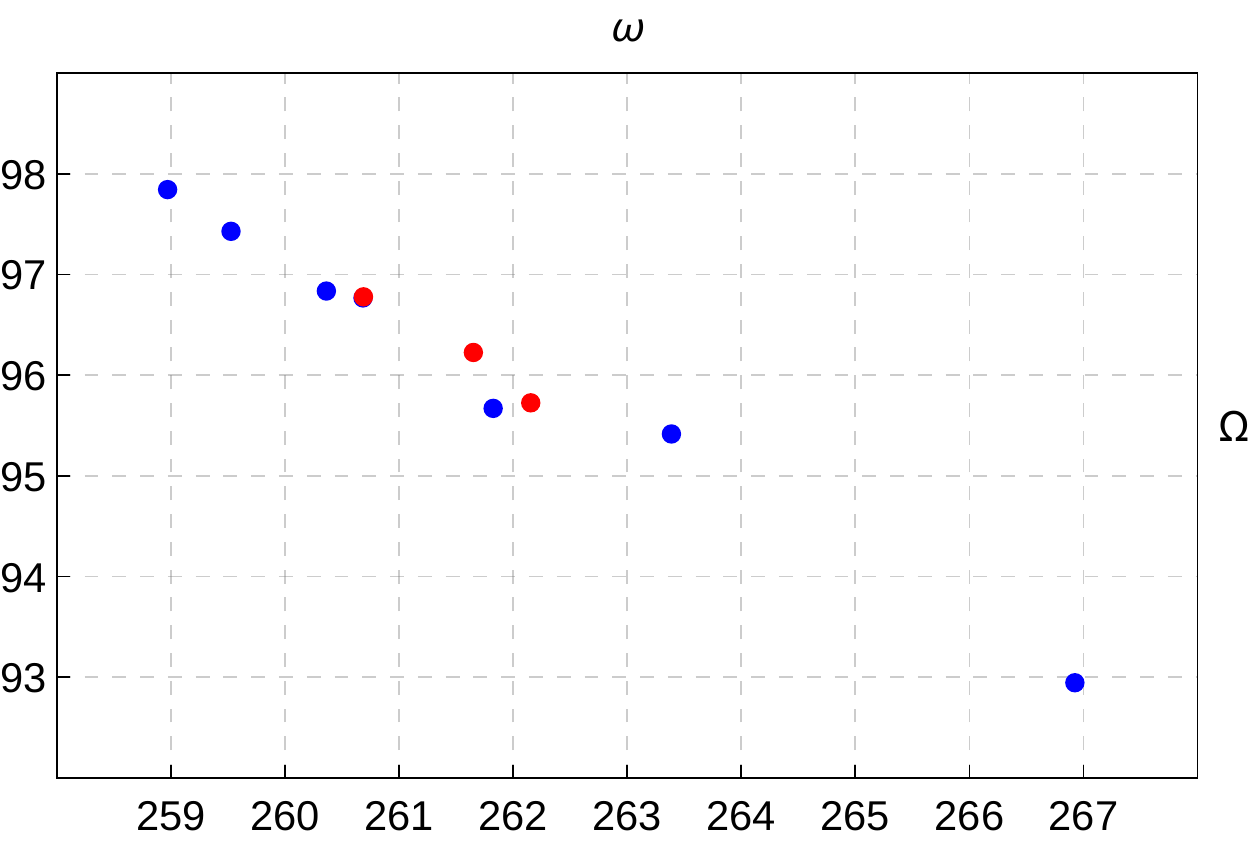}
     \caption{The positions of the Datura family members in $\Omega, \omega$ coordinates. The new members are shown as red dots and the original members as blue dots.}
\label{position}
 \end{figure}

An additional argument which favours these new members belonging to the Datura family, is the calculation of the average orbital elements over the last 800~kyrs. Through numerical integration, we made a linear approximation of node longitudes. For 2002~VR17 (338309) and 2002~RH291, we have a very close comparison  in all the orbital elements. Proper elements, calculated by \cite{2003A&A...403.1165K} confirm that 2002~VR17 (338309) and 2002~RH291  belong to the Datura family (Tab.~\ref{synteticpropelements}). In particular, we can point out a very close match of our calculations of averaged semi-major axis with proper values.

A trait common in the Datura family is a linear dependence between $\Omega$ and $\omega$ (Fig.~\ref{position}). This dependence is unique for the Datura family and remains unfound in other young asteroid families (e.g. 1992YC2, Emilkowalski, etc.{\cite{2006AJ....132.1950N}}), and studying it may be key to successfully reconstructing the dynamic history of the Datura family.

\section{Encounters in the Datura family}

As noted above, if we accept the hypothesis about the origin of all  the members of the Datura family in the same event, the new members of the Datura family's orbital evolution does not conflict with the sugested age of $450\pm50$ kyrs. 

\begin{table*}[t]
\caption{Distances in au between members of the subgroup close to epoch $T=157735$ years ago.}
\begin{center}
\begin{tabular}{l r r r } 
\hline 
 T,yr/d,au & \multicolumn{1}{c}{1270 Datura /} & \multicolumn{1}{c}{1270 Datura /} & \multicolumn{1}{c}{215619 /}  \\
  &  \multicolumn{1}{c}{2003 SQ168(215619)}   & 2001 VN36(89309)  &  \multicolumn{1}{c}{2001 VN36(89309)}\\
157730 & 0.0115 & 0.0160 & 0.0178 \\
157740 & 0.0105 & 0.0207 & 0.0145 \\
\hline 
\end{tabular}
\end{center}
\label{clouseenconter}
\end{table*}

Here we give a few arguments in favour of the subsequent breakup. Numerical integration shows, that two new members of the Datura family have a few recent close encounters with other known members. However, multiple encounters are far more interesting. Below we consider one interesting case of such an event.

A true anomaly of asteroids is most sensitive at the semi-major axis of initial variations. If we speak about a difference of $10^{-6}$ au in the initial semi-major axis, we have a variation in the asteroid position of about $10-15$ degrees after 100 kyrs. In reality, this means that we cannot determine the position of any minor planet along its orbit with the required degree of accuracy after a few thousand years. After such a time interval, we cannot determine the exact longitude of the studied close encounter and minimal distance between the two asteroids. The YORP effect carries a strong influence in the evolution of semi-major axis; the maximum estimated drift rates of kilometer-sized objects being, $da/dt\cong2*10^{-4} au Myr^{-1}$. The mean anomaly can rotate with respect to an orbit with a fixed semi-major axis by 360$^\circ$ in 200~kyrs \citep{2006AJ....132.1950N}.

Another point is, for young compact asteroid families, originating in low-velocity breakups, it is true that two minor planets spent a long time close to one another before impact. So, if we want to determine the epoch of a young asteroid family's origin, we need to look for other criteria. 

As noted above, multiple encounters are far more interesting. Detection of such an event would be direct evidence of a breakup epoch. Here we note a triple close encounter with the participation of 1270~Datura, 2001~VN36(89309) and 2003~SQ168(215619)  (Tab. \ref{clouseenconter}) (Hereinafter referred to as the subgroup). According to data from Tab.~\ref{clouseenconter}, there is a small possibility that asteroid 2003~SQ168 (215619) originated in more recent event\citep{2009A&A...507..495V}. Below we give some new suggestions in favour of this point of view.

\begin{figure}
   \centering 
     \includegraphics[width=8.9cm]{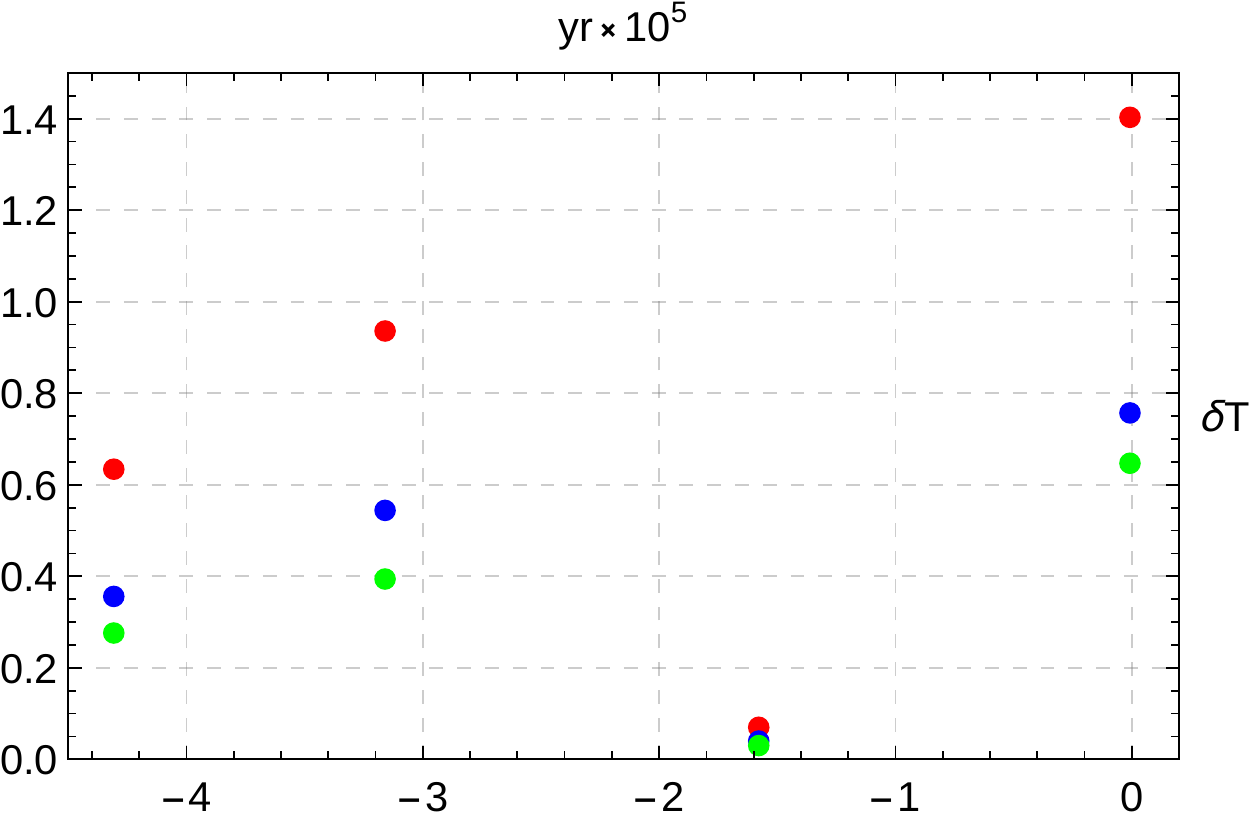}
     \caption{The differences in the period concerning these three bodies is illustrated as follows; red points are for 1270~Datura and 215619 (2003~SQ168/306), blue points for 1270~Datura and 89309 (2001~VN36) and green points for 215619 (2003~SQ168) and 89309 (2001~VN36).}
\label{timediffer}
 \end{figure}

The data in Tab.~\ref{clouseenconter} shows that three members of the Datura family were in very close proximity, with a distance smaller than 0.005 au of their distance from the Sun at an epoch of about $157 735\pm5$ years from now. We can illustrate this fact as follows: 

\begin{equation}
T=\frac{T_{12}+T_{13}+T_{23}}{3} \qquad \delta T_{ij}=T_{ij}-T
\end{equation}
Denote $T_{ij}$ as an epoch of the minimal distance between $i$ and $j$ asteroids, $T$ as the mean epoch of the triple encounter and $\delta T_{ij}$ as the time interval at each partial encounter to $T$. The values of $\delta T_{ij}$ at epoch $157735\pm5$ years ago are by order of magnitude, smaller than other triple close encounters for these asteroids (Fig.~\ref{timediffer}).

The differences in the averaged semi-major axes are determined by averaging their mean motion ie. the period between subsequent close encounters with other minor planets. In our case, this difference is about $5\times10^{-5}$~au. As it seems in Tab.~\ref{time-dependence}, the mean value of such differences in the Datura family is far more than the initial errors in the orbital elements.

\begin{table*}[t]
\caption{Time-dependence of the constant of the subgroup in a 300~kyrs time interval.}
\begin{center}
\begin{tabular}{l r r r } 
\hline 
 Number & \multicolumn{1}{c}{Object} & \multicolumn{1}{c}{Average $a$ [au]} & \multicolumn{1}{c}{Uncertainties
 in $a$}  \\
\hline \hline  
 1270 & Datura  & 2.234694 & 1.1$\times10^{-8}$\\
 89309 & 2001 VN36 & 2.235201 &  7.7$\times10^{-8}$\\
 215619 & 2003 SQ168 & 2.234663 & 2.2$\times10^{-7}$\\
 \hline 
\end{tabular}
\end{center}
\label{time-dependence}
\end{table*}

\begin{figure}
   \centering 
     \includegraphics[width=8.5cm]{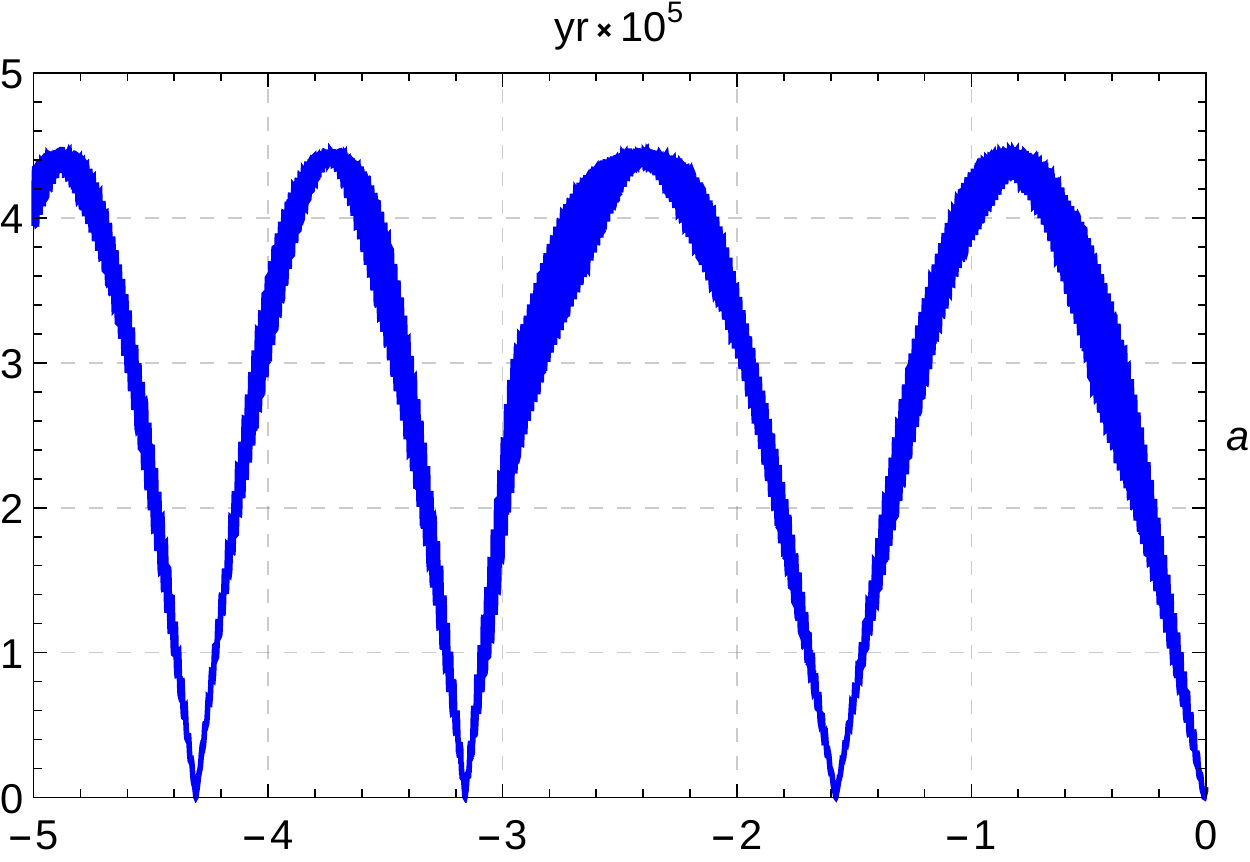}
     \caption{The evolution of  the mutual distance between 1270 Datura and 2003~SQ168(215619).}
\label{Datura-SQ168}
 \end{figure}

The period between two subsequent close encounters between a pair of asteroids in the family can be estimated using:
\begin{equation}
P=\frac{P_1 P_2}{\vert P_1-P_2 \vert}=\frac{{a_1}^{3/2} {a_2}^{3/2}}{\vert {a_1}^{3/2}-{a_2}^{3/2} \vert}
\end{equation}
Calculations based on Tab.~\ref{time-dependence} 2003~SQ168(215619) $P=160.54\pm10$ kyrs is in close agreement with our numerical integrations (Fig.~\ref{Datura-SQ168}). Between 1270~Datura and 2001~VN36 (89309) we have a period of about $10$ kyrs ($9822\pm12$ yrs) again, in close agreement with our numerical integration.

So, we can conclude that the period between two subsequent close encounters between two asteroids of the Datura family is stable and has relatively negligible initial orbit errors. Of course, we cannot say this of the true anomalies or longitudes of the encounter point; they are too sensitive to initial conditions, but we can say this of the epoch of triple encounters with an error margin of about $\pm20$ kyrs, which is comparable with the accuracy of present age Datura family estimations based on angular elements analysis. At least we can detect the time intervals when mutual close encounters of studied asteroids cannot take place. For example, we can completely exclude the time intervals 0--140 kyrs and 180--290 kyrs as a possible origin of 2003~SQ168 (215619).
 
 It is interesting that if our conclusion about a subsequent breakup close to the epoch 157.7~kyrs is true, this means that non-gravitational forces (Yarkovsky  etc.) do not have a significant effect on minor planets in the considered subgroup.  In any case, it is incorrect to exclude 2001~VN36(89309) from Datura family dynamical reconstructions and to reject the possibility of the more recent origin of 2003~SQ168(215619). 

\section{Conclusions}
Three new members of the Datura asteroid family have been found. These new members can help to improve our estimation of the epoch and circumstances of origin. Two of these new members (2002~VR17(338309) and 2002~RH291), have relatively well-determined orbits and belong to the Datura family throughout a time interval of at least 800~kyrs. The third new member, 2014~OE206, has a chaotic resonance orbit like 2001~VN36(89309) and is characterised by a significant increase in semi-major axis about 50~kyrs ago. If we accept the hypothesis about a one-moment origin of the Datura family, the dynamical evolution of 2002~VR17 (338309) and 2002~RH291 does not conflict with the previous estimation of the age of the Datura family ($450\pm50$ kyrs). But studying mutual encounters in this asteroid cluster gives some arguments in favour of a subsequent breakup at the epoch about 157.7$\pm$0.1~kyrs ago. In this more recent event three minor planets 1270~Datura, 2001~VN36(89309) and 2003~SQ168(215619) took part.


\bibliography{mybibfile}

\end{document}